%% file: main.tex
\documentclass[acmsmall]{acmart}

\AtBeginDocument{%
  \providecommand\BibTeX{{%
    \normalfont B\kern-0.5em{\scshape i\kern-0.25em b}\kern-0.8em\TeX}}}

\setcopyright{acmcopyright}
\copyrightyear{2022}
\acmDOI{}

\acmConference[CSCW '25]{CSCW '25: THE 25TH ACM CONFERENCE ON COMPUTER-SUPPORTED COOPERATIVE WORK AND SOCIAL COMPUTING}{November 12--16, 2022}{Taipei, Taiwan}
\acmBooktitle{CSCW '25: THE 25TH ACM CONFERENCE ON COMPUTER-SUPPORTED COOPERATIVE WORK AND SOCIAL COMPUTING,
  November 12--16, 2022, Taipei, Taiwan, Accepted: January 7, 2022}

\usepackage{cleveref}
\usepackage{tikz}
\usepackage[font=small,labelfont=bf]{caption}
\usepackage{hyphenat} 
\usepackage{multirow,color}
\usepackage[super]{nth}
\usepackage{soul}
\usepackage{rotating}
\usetikzlibrary{arrows,positioning,shapes.geometric}
\tikzstyle{startstop} = [rectangle, rounded corners, minimum width=3cm, minimum height=1cm,text centered, draw=black, fill=white]
\tikzstyle{io} = [trapezium, trapezium left angle=70, trapezium right angle=110, minimum width=3cm, minimum height=1cm, text centered, draw=black, fill=white]
\tikzstyle{process} = [rectangle, minimum width=3cm, minimum height=1cm, text centered, draw=black, fill=white]
\tikzstyle{decision} = [diamond, minimum width=3cm, minimum height=1cm, text centered, draw=black, fill=white]
\tikzstyle{arrow} = [thick,->,>=stealth]

\definecolor{dgreen}{RGB}{0, 0, 0}

\begin{document}

\title{Automated Detection of Doxing on Twitter}

\author{Younes Karimi}
\email{younes@psu.edu}
\orcid{1234-5678-9012}
\affiliation{%
  \institution{
  Pennsylvania State University}
  \city{University Park}
  \state{Pennsylvania}
  \country{USA}
  \postcode{43017-6221}
}

\author{Anna Squicciarini}
\affiliation{%
  \institution{
  Pennsylvania State University}
  \city{University Park}
  \state{Pennsylvania}
  \country{USA}}
\email{acs20@psu.edu}
\orcid{0000-0002-7396-1895}

\author{Shomir Wilson}
\email{shomir@psu.edu}
\orcid{0000-0003-1235-3754}
\affiliation{%
  \institution{
  Pennsylvania State University}
  \city{University Park}
  \state{Pennsylvania}
  \country{USA}
}






\renewcommand{\shortauthors}{Younes Karimi, Anna Squicciarini, and Shomir Wilson}

\begin{abstract}
  \input{abstract}
\end{abstract}



\maketitle

\section{Introduction}
\input{intro}

{\color{dgreen}\section{Background and Related Work}}
\input{related-work}

\section{Building a Doxing Dataset}
\input{dataset}

{\color{dgreen}\section{Empirical Analysis of Doxing}}\label{sec:intentions}
\input{intentions}

\section{Automated Detection of Doxing}\label{detection}
\input{detection}

\section{Analyses and Results}
\input{analyses}

{\color{dgreen}\section{Discussion and Limitations}}
\input{limitations}

\section{Conclusion}
\input{conclusion}

\bibliographystyle{ACM-Reference-Format}
\bibliography{z-bibliography}

\vfill
\section*{Appendix}
\input{appendix}

\end{document}

%% file: abstract.tex
\textit{Doxing} refers to the practice of disclosing sensitive personal information about a person without their consent. This form of cyberbullying is an unpleasant and sometimes dangerous phenomenon for online social networks. Although prior work exists on automated identification of other types of cyberbullying, a need exists for methods capable of detecting doxing on Twitter specifically. We propose and evaluate a set of approaches for automatically detecting second- and third-party disclosures on Twitter of sensitive private information, a subset of which constitutes doxing. We summarize our findings of common intentions behind doxing episodes and compare nine different approaches for automated detection based on string-matching and one-hot encoded heuristics, as well as word and contextualized string embedding representations of tweets. We identify an approach providing 96.86\% accuracy and 97.37\% recall using contextualized string embeddings and conclude by discussing the practicality of our proposed methods.

%% file: intro.tex
Doxing is a harmful phenomenon in online text-sharing platforms such as \textit{Pastebin} and online social networks such as Twitter. The term ``dox'' is an abbreviation for ``documents,'' and doxing is the act of disclosing private, sensitive, or personally identifiable information about a person without their consent. Sensitive information can be considered as any type of confidential information or any information that can be used to identify a person uniquely. This information is called \textit{doxed information} and includes \textit{demographic information}~\cite{waseem2016hateful} such as birthday, sexual orientation, race, ethnicity, and religion, or \textit{location information} which can be used to precisely or approximately locate a person such as the street address, ZIP code, IP address, and GPS coordinates. Other categories of doxed information are \textit{identity documents} like passport number and social security number, \textit{contact information} like phone number and email address, \textit{financial information} such as credit card and bank account details, or \textit{sign-in credentials} such as usernames and passwords~\cite{chen2018doxing}. Such disclosure may have various consequences. It may encourage forms of bigotry and hate groups, encourage human or child trafficking and endanger people's lives or reputations, scare and intimidate people by \textit{swatting}\footnote{SWAT stands for Special Weapons And Tactics which is a law enforcement unit that uses specialized or military equipment to respond violent situations} or pushing activists offline, or spread disinformation and rumors about celebrities and public figures~\cite{douglas2016doxing, parab2019twitter}. More importantly, cyberbullying victims may not even recognize such harassment as bullying and may not report or look for help~\cite{van2020multi}.

The state-of-the-art research has focused on different types of cyberbullying and either aimed to identify cyberbullying traces~\cite{agrawal2018deep} or has taken further steps towards differentiating between various categories of cyberbullying. These include but are not limited to \textit{flaming, online harassment, shaming and denigration, impersonation, outing and trickery, exclusion or ostracism, and happy slapping}~\cite{basak2019online, badjatiya2017deep, waseem2016hateful}. 
Cases of doxing or other cyberbullying events may endanger the professional or personal status of victims. The subjects may lose their jobs, families, or even their lives or become excluded from communities. Moreover, once users disclose private information, they cannot withdraw it ~\cite{wang2019donttweetthis, geetha2020will, canfora2018nlp}. Some studies have focused on doxing targets, victims' relationships with doxers, types of doxed information, and common intentions behind doxing. Surprisingly, people not only may dox strangers or those that they dislike but also they may dox those who are known to themselves or even people they like~\cite{chen2018doxing, chen2019doxing}. In another scenario, people may dox those who are harassing them or spreading hate, sexism, or racism to protect themselves or their values~\cite{douglas2016doxing}. 

Social media platforms such as Twitter have taken measures to remove doxing and restrict doxers, primarily using user reports or basic detection methods. Nevertheless, this is non-scalable. Automated detection of doxing episodes may be effective in identifying doxing tweets, raising an alert, and removing the doxed information immediately to mitigate its publicization and dissemination~\cite{song2018personal, khazaei2016detecting, sun2019research, daehnhardt2015usage}. While there are several automated approaches for detecting cyberbullying episodes, to the best of our knowledge, the current methods have focused on cyberbullying detection, regardless of the nature of the attack,  and do not seek to identify tweets that contain doxed information. There are also prior studies on the detection of sensitive information disclosure. Still, they do not differentiate between types of information or use category-based approaches to identify specific categories of private information. Therefore, they are not generalizable enough or capable of determining honest disclosures of sensitive data from malicious disclosures or doxing. 

To study the feasibility of automated doxing detection, we have collected 179,350 tweets that include indicators of potential doxing attacks. We used keywords for two subcategories of sensitive information, {\color{dgreen}Social Security Number\footnote{A nine-digit number issued by social security administration to U.S. citizens, permanent and temporary residents. It is a de facto national identification number and a sensitive identity used for financial and health purposes.} (SSN)} and IP address, to collect the tweets over six months. We have focused on these two types of private information because they are significant. We assumed they are more likely to be misused by identity thieves or cause physical threats to the information owners by revealing their locations. We have also manually labeled 3,131 filtered tweets that contain at least a social security number or an IP address. Based on the collected tweets, we categorize sensitive information disclosure events into three types: (1) purported sharing of first-party sensitive personal information (self-disclosure), (2)  purported malicious sharing of second- and third-party sensitive personal information, and (3) doxing, and propose an automated approach to detect the last two. As we will show in Table~\ref{table:intentions}, the main difference between the last two types is that not every doxing episode is malicious. We have observed scenarios in which a person discloses sensitive information (e.g., about their lost daughter or a missed payment) to seek help or because they do not know that retweeting doxed information is also harmful and helps spread that. Besides these sensitive disclosures, we found harmless or nonsensitive disclosures as information disclosures in which obviously-invalid information (such as 111-11-1111 as someone's SSN) is used or the tweets that contain discussion of sensitive information and corresponding keywords but do not really contain such sensitive information.

We perform a binary classification to exclude self-disclosures and those disclosures that do not target any specific identity and therefore are not sensitive. To this end, we propose nine methods using a suite of different features and feature extraction techniques. We implement a baseline approach using heuristics-based string-matching and compare that with the more robust approaches that use the word and contextualized string embeddings for feature extraction. We present the evaluation metrics of our approaches, the best of which can detect second- and third-party sensitive information disclosures with 96.86\% accuracy, 98.16\% precision, 97.37 \% recall, and 97.76\% F1 score. We have compiled a comprehensive set of potential intentions (acknowledging actual intent is unavailable to us) behind sensitive information disclosures and their consequences through manual inspection of our tweet collection. In summary, our contribution in this paper is three-fold:

\begin{itemize}
\item Create and curate a dataset of 179,350 tweets that are likely to contain doxing episodes, and manually label 3,131 tweets.
\item Present a comprehensive insight into potential intentions behind sensitive information disclosures and differentiate between these motivations in terms of whether they constitute doxing or malicious disclosures and if they are defensive acts.
\item Describe an automated approach for detection of doxing and malicious sensitive information disclosures about second- and third-parties{\color{dgreen}.}
\end{itemize}

%% file: related-work.tex
\subsection{Definition of Doxing}
The term \textit{doxing} comes from the phrase ``dropping documents'' or ``dropping dox,'' and ``dox'' is an abbreviation for ``documents''~\cite{douglas2016doxing}. Doxing can be referred to as a type of cyberbullying~\cite{chan2019child} and an act of collecting private, sensitive, or personally identifiable information and documents about others and publicly disclosing them without their consent. This information may be publicly available (e.g., due to data breaches in organizations or agencies) but hard-to-access or distributed between various sources~\cite{song2018personal}. Moreover, there are scenarios in which the doxer obtains sensitive information from the victim willingly or unwillingly~\cite{douglas2016doxing}. The former may happen in the form of \textit{outing} where someone shares private information intended to be shared only with them or with a group of friends, such as text messages or private photos and videos. Still,  they re-share and expose that private information to an unwanted or unforeseen audience. The latter is when someone elicit private information by \textit{trickery}, and in various forms such as \textit{social engineering}~\cite{dalton2019modeling}, \textit{phishing}, \textit{spear-phishing}~\cite{stringhini2015ain}, or \textit{impersonation}~\cite{dalton-etal-2020-active}. Consequences of doxing may range greatly. {\color{dgreen}Although doxing may lead to criminal charges such as cyberstalking and harassment, doxers are often hard to hold accountable, especially if they act anonymously~\cite{GamerGateThreats}. Examples of severe cases for which consequences are known include online and offline harassment of women in the gaming industry. Other instances are swatting and prank calls to send police to a journalist's or politician's home and investigate a false emergency report, rape, bomb, and death threats~\cite{ScaryDoxing}, and even death~\cite{swatting-death}}.\\

Although prior studies do not propose any automated approach for detecting doxing episodes and malicious disclosures of private information on Twitter specifically, these incidents are still considered cyberbullying and online harassment and consist of personal information disclosures. While cyberbullying events may include doxing and disclosures of victims' private information~\cite{chan2019child}, they do not have to. Based on the existing cyberbullying detection methods, sensitive information disclosure incidents discussed in this paper are very likely to be missed. To the best of our knowledge, existing work has focused on various types of cyberbullying, their identification, private information detection, and social aspects of doxing. But, an automated approach is still required to detect doxing and malicious disclosures and mitigate their harms and consequences on Twitter.\\

\subsection{Cyberbullying, Doxing and Social Aspects}
Cyberbullying~\cite{bellmore2015five} is a subset of bullying and consists of any type of online harassment performed using communication technologies by a user or a group of users against other users~\cite{al2016cybercrime}. The significant differences between traditional bullying and cyberbullying are that (1) the victims are more reachable for cyberbullying regardless of their location and (2) its consequences can be more persistent~\cite{xu2012learning, bellmore2015five}. In addition to Twitter, previous research has studied instances of online harassment in other online social networks and forums such as YouTube, Instagram, MySpace, Wikipedia, Slashdot, and Kongregate~\cite{van2020multi}. Ioannou et al. have surveyed research designs in the cyberbullying realm and the associated qualitative, quantitative, and experimental studies. They have also analyzed automated cyberbullying detection approaches and concluded that the current algorithms are capable of detecting cyberbullying only up to some degree of success and urge further investigations along with recommending future research paths~\cite{ioannou2018risk}. They additionally provide a typology of cyberbullying actors. We also have analyzed the user accounts behind our doxing dataset and compared their profile and network characteristics in Table~\ref{table:individual-attributes}.

Douglas has presented a conceptual analysis of doxing, some of its types and motivations, and discussed whether, in which contexts, and to what extent, doxing might be justified~\cite{douglas2016doxing}. His work also introduces three types of doxing, but we will expand those types and propose a more comprehensive set of potential motivations and types of sensitive information disclosure.
In a study of high school students in Hong Kong, Chen et al. have analyzed doxed information, doxing targets and platforms, and their deviations based on the gender~\cite{chen2018doxing, chen2019doxing}. They have also studied the social and psychological impacts of doxing incidents on the victims. Moreover, Lozano-Blasco et al. have studied the duality of being a cybervictim and a cyberbully and discussed how these actors may suffer from anxiety and depression~\cite{lozano2020being}.

Snyder et al. have proposed an approach for detection of doxed information on text sharing platforms such as \textit{pastebin.com} and \textit{4chan.org}~\cite{snyder2017fifteen}. Doxing on a social networking platform such as Twitter represents a different set of challenges because of the availability and rapid distribution of exposed information. For instance, if someone's SSN is replied to their tweet, that information immediately becomes available to all the author's followers and the victim. Conversely, people will have to search for specific individuals on the aforementioned websites to see whether their private information has been shared and is still available there. The authors have also introduced four types of motivations for doxing and manually assigned a portion of their doxing samples to each of these motivations. Although the presented motivations can be observed in our dataset, they cannot properly explain and differentiate all our disclosure incidents. 
While their main goal is to automatically identify doxing samples in text-sharing websites, we focus on identifying various types of information disclosure and developing an approach for detecting disclosures of two categories of sensitive information that can be more harmful.

\subsection{Private Information Detection}
Confora et al. have created 97 heuristics based on recurrent patterns to identify Location and Emotion information disclosures~\cite{canfora2018nlp}.
Mehdy et al. have proposed a sentiment-aware privacy disclosure detection framework based on neural networks~\cite{mehdy2020user}. Deodhar et al. proposed an approach for identifying private information and its category~\cite{deodhar2017analysis}. Their approach includes extracting 105 topics and 50 features based on the dependency graph structures of tweets. Other features include location information (if the tweet is geotagged) and user-mentions (if the author \textit{@mentions} another user). Although topics have been extensively used for text classification in various domains~\cite{caliskan2014privacy}, our proposed approach does not rely on them. This is because our goal is to provide a generalizable method for automated detection of different categories of sensitive information, disregarding category-specific topics.
Moreover, based on intuition obtained by manual inspection of our tweet collection, we do not believe in topics to be distinguishing enough between our target tweets. Dependency graph structures have also been used in various problems such as syntactically-correct sentence creation for generating memorable passwords~\cite{7736457}. However, the authors of~\cite{deodhar2017analysis} had to limit the number of dependency graph features to 50 and remove redundant sub-trees containing proper nouns to reduce the variety of words and relationships they may get. While they illustrate a high performance on differentiating between categories, their private information detector yields only 80.5\% accuracy. Additionally, not all privacy leakages are malicious or can be considered doxing. We will show in~\S\ref{sec:intentions} what other scenarios are in which private information may get disclosed, intentionally or unintentionally, maliciously or unmaliciously. Caliskan-Islam et al. have analyzed users' privacy behavior on Twitter and proposed an approach for scoring users based on private information disclosure in their tweets~\cite{caliskan2014privacy}. They detect the presence of private information and calculate privacy scores for users' timelines.\\

Table~\ref{table:similar-works} compares our proposed approach with closest similar studies. Specifically, none of these approaches is capable of detecting second- and third-party sensitive information disclosures on Twitter and differentiating them from self-disclosures and nonsensitive disclosures. The first study aims to find sensitive location and emotion information~\cite{canfora2018nlp}, and the second one seeks private information in general~\cite{deodhar2017analysis} and disregards the notion of cyberbullying. We will elaborate on our own method, and the significance of our detection capability (i.e., differentiation of second- and third-party disclosures from nonsensitive- and self-disclosures) in \S \ref{string-embedding} and \S \ref{sec:intentions} respectively.

\begin{table*}[ht]
\centering
\begin{tabular}{|c|c|c|} 
 \hline
 Method/Features & Detection Capability & NS/\nth{1} vs \nth{2}/\nth{3} Party \\ [0.5ex] 
 \hline\hline
 Syntactic Heuristics~\cite{canfora2018nlp} & Sensitive \textit{location} and \textit{emotion} & Does not differentiate\\& information disclosure &\\\hline
 Topics + Grammar & Private information & Does not differentiate\\Dependency~\cite{deodhar2017analysis} & & \\\hline
 Contextualized String & Doxing and malicious sensitive & Differentiates\\ Embedding (present work) & information disclosure & disclosures\\ [0.25ex]\hline
\end{tabular}
\caption{Comparison of similar works, their methods and detection capabilities. ``NS'' stands for ``Nonsensitive'' disclosures. None of the earlier works is capable of differentiating nonsensitive and self-disclosures (\nth{1} party) from doxing and malicious disclosures of second- and third-parties' sensitive information.}
\label{table:similar-works}
\end{table*}

%% file: dataset.tex
Although sensitive information disclosure, doxing, and other sorts of cyberbullying can be seen extensively in different platforms such as forums~\cite{mehdy2020user}, emails, and online social networks~\cite{chen2019doxing}, our focus in this paper is on tweets and Twitter data. Such an analysis requires a tremendous amount of data~\cite{van2020multi} because our desired samples that contain sensitive information and doxing episodes constitute a small portion of all the posts on an online social network. We focus on Twitter because it is a straightforward source of large volumes of online social networks data and provides a public REST API for collecting the streams of tweets{\color{dgreen}.}\footnote{https://developer.twitter.com/en/docs/tutorials/consuming-streaming-data} While some datasets are available for detecting other types of cyberbullying on Twitter~\cite{bellmore2015five, resnik2016celebrities}, they have not focused on doxing and sensitive information disclosure specifically. Therefore, they mostly do not contain any sensitive information, and we had to collect a new set of tweets. We cannot publicly release our collected tweets to preserve users' privacy. However, we can provide the tweet IDs to the research community upon request.

\subsection{Collecting Potentially-Sensitive Tweets}
The streaming API provides several capabilities, such as collecting tweets based on keywords as initial filters for discarding obviously-not-related tweets. These include tweets that do not contain any sensitive information or are not talking about any of the categories we presented earlier: social security number and IP address. Researchers have used examples of such filters and keywords as simple-but-effective filtering of unrelated tweets. The keywords like ``\textit{bully},'' ``\textit{bullied},'' and ``\textit{bullying}'' are used by Xu et al. \cite{xu2012learning} to filter out tweets that are less likely to contain cyberbullying traces. The API also lets us excludes tweets that are written in other languages and only collect English tweets{\color{dgreen}.}\footnote{https://developer.twitter.com/en/docs/twitter-api/getting-started/guide}

We used the Twitter streaming API and Tweepy{\color{dgreen},}\footnote{http://docs.tweepy.org/en/v3.4.0/streaming\_how\_to.html} a Python wrapper library~\cite{deodhar2017analysis}, to collect {\color{dgreen}public} tweets. As shown in Table~\ref{table:keywords}, for each category of sensitive information, {\color{dgreen}we collected the corresponding tweets using related keywords that were potentially indicative of identifying or location information (i.e., SSN and IP address).} Our collection consists of tweets published from June 15 to December 17, 2020. {\color{dgreen}Note that since our goal was to collect a sample of isolated tweets that contains doxing incidents, the collection of data was not intended to be contiguous.} Additionally, we collected hundreds of thousands of tweets during this period using more keywords to perform an initial analysis on other subcategories of sensitive information as well. Table~\ref{table:keywords} presents the total number of tweets we collected for each subcategory and the number of tweets remaining after initial structural filtering.

It is worth noting that although we observed remarkable cases where different categories of sensitive information were shared  in a single tweet, for our analyses in this paper, we only used and presented the number of tweets containing each subcategory of sensitive information solely based on the tweets we collected with the corresponding set of keywords. That is, if there is a tweet in the SSN collection that contains an IP address, we did not consider that in the IP address collection. This is because we wanted to present a more clear perspective toward the ratio and frequency of actual sensitive information once we use the corresponding keywords. Moreover, as we will discuss in \S~\ref{sec:intentions}, in some cases, such as a report of a doxing episode, the doxed information may be found in the quoted status, not the actual tweet. In such cases, the corresponding keywords might also be found in the quoted tweet, but Twitter streaming API still collects that tweet. Regardless, we analyze the whole tweet as a concatenation of the two tweets. Furthermore, we did not collect tweets of each subcategory continuously. Some tweets may be retweeted often or get quoted several times, affecting the number of unique samples in our stream. Therefore, after collecting the first batch of  SSN tweets, we moved to a different subcategory.

\begin{table*}[ht]
\centering
\begin{tabular}{|c||c|c|c|c|c|c|} 
 \hline
 Category & Sub- & Keywords & Tweets & Filtered & \nth{2}/\nth{3} & SD/NS\\&Category&&& {\color{dgreen}\& Labeled} & Party &\\ [0.5ex] 
 \hline\hline
  \rule{0pt}{1ex} Identity & & ssn, ssa, &&&&\\Docs & SSN & social security number, & 140,510 & 520 & 219 & 301\\ && social security administration &&&&\\\hline 
 Location & IP & ip address & 38,840 & 2,611 & {\color{dgreen}1916} & {\color{dgreen}695}\\Info & Address &&&&&\\
 [0.25ex]\hline
\end{tabular}
\caption{Sensitive information categories, corresponding keywords and collected tweets. ``SD'' and ``NS'' stand for ``Self-Disclosure'' and ``Nonsensitive'' respectively. Additionally, ``\nth{2}/\nth{3} Party'' tweets are those that disclose sensitive information about others.}
\label{table:keywords}
\end{table*}

\subsection{Initial Structural Filtering and Preprocessing}\label{sec:init-filtering}
Keywords are very effective in filtering out unrelated tweets that are not about any sensitive topic, such as a tweet that reports the scores of a football game. But, even after such filtering, there exist plenty of tweets that may include sensitive keywords, such as social security number, but do not contain that kind of information. An example of this is when someone tweets about a new procedure in the social security administration for issuing SSNs. Our goal in this paper is to detect the disclosure of sensitive information, while these tweets may not actually have any SSNs. To remove these, we have implemented separate filters for detection, and basic validation of SSNs and IP addresses according to their standard structures.

According to the Social Security Administration{\color{dgreen},}\footnote{https://www.ssa.gov/history/ssn/geocard.html} each SSN consists of three parts of \textit{area}, \textit{group}, and \textit{serial} numbers. The first two used to indicate different areas in the states. But, since SSN Randomization{\color{dgreen},}\footnote{https://www.ssa.gov/employer/randomization.html} these number are being generated randomly. The current restrictions are either the syntax which includes three parts of three, two, and four digits, or area number exclusions. Area numbers cannot be 666 or any number between 900 and 999. Besides, none of the three segments can be all zero. Therefore, we used regular expressions to exclude the invalid SSNs.

For IP address, this filter includes the syntactic structure of IPv4 in which none of the four digits in each IP segment can go beyond 255. Moreover, we removed trivial invalid IP addresses such as ``$0.0.0.0$'' or ``$8.8.8.8$,'' or the ones starting with ``$192.168$'' or ``$127.0.0$,'' and those that have digits higher than 255 for each IP segment. Only the digit validation filter removed 329 invalid IP addresses, and in total, we filtered out 36,229 tweets that were collected using our keyword for the IP address subcategory but did not really have any IPv4-structured address, or their addresses were obviously invalid or nonsensitive. Finally, the labels ``\nth{2}/\nth{3} Party'' and ``Nonsensitive'' in Table~\ref{table:keywords} refer to those tweets that we believe respectively, do or do not contain sensitive information disclosures about second- or third-parties.

We removed redundant characters, stop and functional words as they are the main parts of tweets that can be eliminated to enhance the quality of tweets while keeping the most essential and distinguishing parts. Although such preprocessing of textual data has been studied and extensively used in natural language processing tasks~\cite{xu2019detecting}, we conducted several experiments with and without performing the preprocessing to identify its effects on the performance of our approach.

\subsection{Labeling}
To perform a supervised automated classification and differentiate doxing and malicious private information disclosures from non-doxing tweets and self-disclosures~\cite{umar2019detection}, a representative sample of tweets needs to be manually labeled by domain experts. Prior studies have used a comparable dataset size (4,865 tweets including 91 tweets labeled as cyberbullying~\cite{huang2014cyber}, or 1,762 tweets consisting of 684 cyberbullying incidents~\cite{xu2012learning}), to perform automated detection of cyberbullying. A more recent study of cyberbullying detection~\cite{al2016cybercrime} has used 10,007 manually-labeled tweets, 599 of which contain indications of cyberbullying behaviors. We collected and labeled the tweets reported in Table~\ref{table:keywords} in multiple rounds until we collected enough positive and negative samples for each subcategory of sensitive information according to the previous studies. {\color{dgreen}The first author, a doctoral candidate with research experience in the areas of cyberbullying and social media analysis,} performed the labeling task. 

To validate the labeling and underlying decision-making process, we asked two other researchers experienced in areas of privacy, cybersecurity, natural language processing, and social network analyses to label a small sample of {\color{dgreen}200} tweets. {\color{dgreen} We used this sample for measuring the inter-rater reliability and only used the first author's annotations for all other analyses. This sample was selected from our collection such that each category of sensitive information contributed to a half of the sample. We also removed the retweets and selected tweets from each category such that they have the least similarity with other tweets in their writing and vocabulary. By doing so, we could provide a more diverse sample of tweets to our annotators. We removed the handles ($@username$), tokenized the tweets, and obtained a list of \textit{tokens}.\footnote{A token is an instance of a sequence of characters in some particular document that are grouped together as a useful semantic unit for processing. It can be loosely referred to as ``terms'' or ``words'': https://nlp.stanford.edu/IR-book/html/htmledition/tokenization-1.html} Before tokenization, we removed non-alphabetical characters (including all digits and excluding white spaces between the tokens) from the tweet texts. This is done because our goal was to compare the similarity between the vocabulary and the actual content of tweets and ignore the subtle differences between them that are due to the presence of different Twitter handles, SSNs, IP addresses, or emoticons. 

We combined all the tokens extracted from the tweets and created a bag of 4,363 unique tokens representing the vocabulary space of our tweets. We assigned feature vectors of this size to each tweet by counting the number of tokens from this bag that were present in the corresponding tweet. This resulted in 356 and 1,165 sparse vectors of 4,363 elements for SSN and IP address tweets, respectively (out of the totals of 520 and 2,611 for SSN and IP address that include retweets). We calculated pairwise cosine similarity~\cite{HAN201239} between every pair of vectors in each category to determine token-level similarities amongst the tweets~\cite{almuhimedi2013tweets}. This yielded 356 matrices of size $356 \times 356$ and 1,165 matrices of size $1,165 \times 1,165$ (each element represents the similarity of the corresponding tweets in that category). Using these matrices, we computed the sum of all the similarities in each tweet's matrix to identify its overall similarity to the other tweets in our dataset. We selected 100 tweets from each category (200 tweets in total) that had the least overall similarities to other tweets in the same category.}

{\color{dgreen}The annotators were given a short set of instructions explaining our definition of doxing and malicious disclosures and how to annotate the tweets. We created a table mentioning the most common indicators for each of the two classes so that the annotators can easier and more consistently follow the instructions through the whole process.\footnote{Annotators' instructions are reported in the Appendix (\S\ref{sec:appendix})}} We calculated {\color{dgreen}Fleiss' Kappa coefficient~\cite{fleiss1971measuring}} to measure agreement between our annotators {\color{dgreen} and obtained the value of 0.50 between all three annotators which is considered as ``moderate agreement'' by Landis and Koch's benchmark scale~\cite{landis1977measurement} and ``intermediate to good agreement'' by Fleiss's benchmark scale~\cite{gwet2014handbook}. We also calculated the pairwise Cohen's Kappa coefficient~\cite{kraemer2014kappa} amongst the three annotators and obtained an identical mean value. We argue that some of} these differences are because not only the main annotator was a domain expert, but also he had seen a lot more samples and obtained more contextual information about similar tweets after multiple rounds of labeling. Moreover, we manually inspected instances of disagreement amongst the annotators. We observed some tweets that seemed to be disclosing sensitive information but were not directly associated with specific and unique identities and were mistakenly labeled as doxing by other annotators.

%% file: intentions.tex
\label{sec:intentions}
\subsection{Potential Motivations and Consequences of Doxing}

We differentiate between purported and malicious purported sensitive information disclosure and doxing. There can be scenarios where the disclosure is doxing, but it is not necessarily malicious and vice versa. An example of the former is when a parent is desperately looking for his lost daughter and benignly shares her private information with the hope of paving the path for her to be found. We consider all these disclosures as purported, claimed, or outwardly appearing to be sensitive information. This is because we cannot verify any of this information with their claimed owners, barring some clear spamming messages or fake information.
Further, we consider self-disclosure of private information as non-doxing and non-malicious. Self-disclosure is defined as the act of someone publicly sharing his private information. This disclosed information can be either real or fake. 
However, in a different scenario, one may masquerade as a known character and disclose her private information using the forged account so that others think she has exposed this information herself. As it might be much more difficult to identify such masquerading even manually and it requires further information about the incident and the real identity, our approach considers this case as self-disclosure.


A foundational study of doxing by Douglas categorized doxing episodes into three coarse-grained types of \textit{deanonymization}, \textit{targeting}, and \textit{delegitimization}, and considered \textit{exposing wrongdoing, humiliating, threatening,} and \textit{punishing} as some motivations behind doxing. These three categories are consecutively defined as releasing information that reveals the identity of the person who has previously been anonymous, revealing information about an individual that allows her to be physically located, and releasing private information to undermine the owner's credibility or reputation~\cite{douglas2016doxing}. However, based on our collection of tweets and the doxing samples that we identified, we believe these categories are not representative enough for distinguishing among doxing types. In Table~\ref{table:intentions} we introduce a more comprehensive set of common motivations for disclosure of sensitive information based on our observations. Note that some disclosure incidents may fall into multiple categories and consist of multiple intentions and consequences. As it is shown, all of these motivations can lead to a doxing episode (depending on the situation) except the last row in which people expose their own private information. Moreover, ``doxing report'' and ``help-seeking'' may or may not be malicious, but because they promote the exposed sensitive information and help that spread more, we consider them doxing. Additionally, Table~\ref{table:intentions} illustrates whether an act of sensitive information disclosure may aim to defend the actor and their values. For instance, a victim of cyberbullying or misogyny may disclose the attacker's sensitive information in a defensive, malicious or unmalicious reaction to scare the subject or get help from others.\\

\paragraph{\textbf{Endanger}}
Disclosing some private information such as location information that can facilitate locating people, or sexual orientations, may put people or their children at the risk of physical threats, human trafficking, reputational risk, sexualized misrepresentation, or hypocrisy~\cite{douglas2016doxing}. We observed 654 tweets revealing IP addresses alongside GPS coordinates, some of which claimed to be related to specific users. While we cannot verify or refute this information, we found similarities and discrepancies between the IP addresses' approximate locations and the reported coordinates after several IP address lookups. Some discrepancies can be explained if the IP address is for the user's {\color{dgreen}Internet} service provider or if she has used a VPN. So, those coordinates that really belong to the purported users can help locate them more precisely. 

\paragraph{\textbf{Scare, distress or panic}}
Examples include pushing journalists and political activists offline, alarming, intimidating, or blackmailing people such as women in the gaming industry in the ``GamerGate'' harassment campaign, and threatening them to rape or death~\cite{douglas2016doxing, waseem2016hateful}. A different example is \textit{swatting} in which an attacker makes a hoax call to the police and claims that there is a violent perturbation at the victim's location, demanding an armed police force~\cite{douglas2016doxing}. While we cannot verify whether the purported information really belongs to the victims, in a different distressing scenario, a malicious user may reply a random IP address and GPS coordinates to a tweet of a non-technical user who does not know how to obtain this information about themselves.

\paragraph{\textbf{Defame or denigrate}} 
Common cyberbullying events on social media websites may contain direct attacks, including but not limited to shaming and denigration in which social media users troll individuals in order to silence, publicly embarrass, or discredit them. These attacks that mostly can be seen against public figures, celebrities, and politicians may occur along with disclosing their real or fake private information, spreading rumors and misinformation about them. An example of this is:
\begin{quote}{\color{dgreen}
    ``\textit{@\texttt{[USERNAME1]} @\texttt{[USERNAME2]} @\texttt{[USERNAME3]} \texttt{[FNAME1 LNAME1]}'s Social Security number is $***-**-****$ and was issued by \texttt{[STATE]}, though he never lived in the state. In fact, that number was actually issued to \texttt{[FNAME2 LNAME2]}, who was born in \texttt{[YEAR]}!}''}
\end{quote}
The currently-suspended author had repeatedly tweeted that SSN and mentioned different people. 

\paragraph{\textbf{Digital vigilantism}}
Doxing is sometimes used as a tool to expose wrongdoing~\cite{douglas2016doxing} and get the targets fired from their jobs, shamed in front of their neighbors, run out of town, or encourage reform or remorse of hate groups, or against discrimination, racism, sexism~\cite{badjatiya2017deep, waseem2016hateful}, and homophobic people. An example of this type is \textit{human hunting} or the \textit{human flesh search engine}. Chinese media originally used this term to refer to the practice of searching for people (mostly government officials) online and hunting down those who are attracted to the users' wrath~\cite{gao2014hunting}. 
Although such behaviors may be against the law or unethical, as it is typically performed by self-appointed groups without legal authorities, they may not be fair or may expose the victim's private information more than the extent that is required to expose his wrongdoing~\cite{solove2007future, douglas2016doxing}.

\paragraph{\textbf{Doxing report}}
Doxing report is when \textit{Eve} shares private information about \textit{Alice}, and \textit{Bob} re-shares that. Besides the initial doxing that was conducted by \textit{Eve}, we consider the re-sharing act of \textit{Bob} as doxing too because regardless of having malicious intentions or not, he has helped spreading \textit{Alice}'s private information. But, if \textit{Alice} exposes her own private information (self-disclosure), and \textit{Bob} retweets that, we still consider that as non-doxing.

\paragraph{\textbf{Help seeking}}
We observed a different type of sensitive information disclosure when someone is trying to get help from others. An instance is a parent who shares private information about their lost daughter. A similar example is where someone shares their information such as SSN and phone number and asks for the help of some sort. While such examples may seem like requests for help, based on the type and abundance of information that some of them had disclosed, such as SSN and date of birth, we cannot make a concrete judgment whether they are genuine requests or implicit doxing of the purported lost person. They might even try to connect various information about an identity without explicitly mentioning her name in the tweet.

\paragraph{\textbf{Self-Protection}}
Online users could also dox others in a response to trolls or as a denunciation or self-protection measure against misogynists or racists to stop them or get help from others to find, report, call-flood, or punish the cyberbully. They may also expose information about a scam or an attacker who has targeted themselves to mitigate further frauds.

\paragraph{\textbf{Self-Disclosure}}
This is when someone exposes her own private information to the public: 
\begin{quote}{\color{dgreen}
    ``\textit{@\texttt{[USERNAME]}  my name is \texttt{[FNAME LNAME]}; My social security: $***-**-****$; My phone number: $(***)***-****$; I files for a quit of time now since $**/**/****$  I sent the questionnaire and everything I haven't heard anything yet.}''}
\end{quote}
Some people may also reveal their private information in pursuit of social rewards~\cite{umar2019detection}.
Note that as claims cannot be verified, self-disclosure may only be apparent. To exemplify, we observed a Twitter account in our tweet collection with the screen name of ``{\color{dgreen}\texttt{[FNAME]Top1Global}},'' name of ``{\color{dgreen}\texttt{[FNAME LNAME]}},'' and description of ``\textit{[ World’s Top 1\% and Assets infinitely ] \# 1 First of human Pioneered Entire Digitals Webcams, Hybrids Tech.}'' The tweet text was ``\textit{@USSupremeCourt @wto Plaintiff: Mr. {\color{dgreen}\texttt{[FNAME MNAME-INITIAL LNAME]}}(U.S. \#SSN $***-**-****${\color{dgreen})}\footnote{We have anonymized sensitive information and replaced them with \texttt{``[ALL-CAPITAL WORDS]''} and asterisks to preserve the subject's privacy in case the purported information is real.} World’s Top 1 \% entire future humanity infinitely;}'' While this account claims to belong to {\color{dgreen}\texttt{[FNAME LNAME]}}, a United State's attorney, we believe it is masquerading as him. The reason is that the account has had 37,264 tweets, was following 881 accounts and did not have any followers at the time we collected this tweet. The account description also is full of flaws and does not seem to be written by such an attorney. Moreover, the account is no longer accessible and has been suspended by Twitter. 

\begin{table*}[ht]
\centering
\begin{tabular}{|l||l|c|c|c|}
 \hline
 Intentions & Examples and consequences & Def. & Mal. & Dox.\\ [0.5ex] 
 \hline\hline
 
  & May cause human trafficking, reputational risk, &&&\\Endanger & physical threats, sexualized misrepresentation, or & N & Y & Y\\& hypocrisy &&&\\\hline
 
 Scare, distress,& Swatting or pushing activists offline by intimidating, & N & Y & Y \\or panic & alarming, or blackmailing&&&\\\hline
 
 Defame or & Presenting disinformation, rumors, or real and private &&&\\denigrate &  info about celebrities or public figures to discredit & N & Y & Y | N\\&  them&&&
 \\\hline 
 
  & To get the targets fired from their jobs, shamed in  &&&\\Digital & front of their neighbors, run out of town, or & Y & Y & Y | N \\vigilantism & encourage reform or remorse of hate groups, &&& \\& discrimination, racism, sexism, and homophobia &&&\\\hline       
 
  & Describes, promotes/accuses someone about a doxing &&&\\Doxing report & episode that happened (by quoting, replying, or & Y | N & Y | N & Y\\& indicating)&&&\\\hline 
 
 Help seeking & Disclosing sensitive info about relatives by getting or  & Y | N & Y | N & Y\\ & pretending to be scared or worried about them &&&\\\hline 
 
 Self-Protection& Public denunciations against misogynist trolls & Y & Y | N & Y
 \\\hline 
 
 Self-Disclosure & Revealing the author as the bully, victim, defender, & Y | N & N & N\\& bystander, assistant, or re-enforcer&&&
 \\ [0.25ex]\hline
\end{tabular}
\caption{Common motivations and intentions behind public sensitive information disclosures. ``Y,'' ``N,'' ``Def.,'' ``Mal.,'' and ``Dox.'' stand for ``Yes,'' ``No,'' ``Defensive,'' ``Malicious,'' and ``Doxing'' respectively. We use ``Y | N'' for different situations in which a category may or may not have a specific characteristic.}
\label{table:intentions}
\end{table*}

{\color{dgreen}
\subsection{User Attribute Analysis}
To obtain insights into behavioral differences between the accounts that  tweet doxing, sensitive,  or malicious content,\footnote{For simplicity, we call the two classes ``malicious'' and ``benign'' in this subsection, but we will further differentiate between these four types in \S \ref{detection}.}  we have explored potential correlations between their social network, profile characteristics, and their malicious behavior,  and inspected individual attributes of these Twitter users. As illustrated in Table~\ref{table:individual-attributes}, the number of tweets and unique accounts that belong to the malicious class is higher than the other class. The reason for not using a balanced dataset (concerning either the number of tweets or the number of users) is that we aimed to keep our dataset and experiments as realistic as possible. In practice, once the tweets with potentially sensitive information (SSN and IP address) have passed our initial structural filtering, this can be more similar to their distribution. While we unexpectedly did not observe a drastic distinction in network characteristics of the users in our two classes, some of their profile characteristics have considerable differences. 

Furthermore, a relatively (see the last column) higher number of the benign users in our dataset have a lower number of tweets and liked statuses, have set profile banners or changed their default profile theme. As discussed in prior studies~\cite{douglas2016doxing, ioannou2018risk}, malicious users may tend to remain obscure in online platforms to avoid persecution and punishment or protect their reputation while engaging in doxing or other forms of cyberbullying anonymously. Therefore, it is not surprising to see that a relatively higher number of malicious accounts (0.72 times more) had very short names (e.g., only two characters or an emoji), and 92.81\% of all the accounts in the malicious class have been created very recently (since 2019) and do not have a long history of activities. We also found tweets from a user with the verified badge in our malicious samples who has posted this tweet:

\begin{quote}
``\textit{\texttt{[FNAME]}’s social security number!
$***-**-****$}''    
\end{quote}

In reply to this tweet:

\begin{quote}
``\textit{What can you recite from memory?}''  
\end{quote}

We cannot verify if the SSN is valid and really belongs to the claimed person or who the potential victim is because we only have his first name without any last name or connections to any specific user. But, the potential doxer and his followers or the user who has posted the question might still have some shared knowledge which leads to the identification of the target solely based on his first name. Therefore, we have labeled this tweet as a nonconsensual disclosure of sensitive information (malicious class) where the potential third-party victim may not promptly get notified about such disclosure. We also inspected a few other attributes, such as whether the users have specified their locations or websites in their profiles or set any profile images, but did not observe notable differences among the two classes.}

\begin{table*}[ht]
\centering
\begin{tabular}{|l||c|c|c|} 
 \hline
 \textbf{Attributes} & \textbf{Malicious | Doxing} & \textbf{Benign | Self-Disclosure} & \textbf{Mal./Ben.}\\ [0.25ex] 
 \hline\hline
 Total samples & 2,135 & 996 & 2.14\\
 Unique users & 1,681 & 507 & 3.32\\
 Mean status (tweet) count & 20,928 & 85,625 & 0.24\\
[0.25ex]\hline\hline
\textbf{Network characteristics} &&&\\\hline
No followers & 1.19 \% & 2.56 \% & 0.47\\
No friends (followings) & 0.60 \% & 1.97 \% & 0.31\\\hline
\textbf{Profile characteristics} &&&\\\hline
No Favorites & 0.71 \% & 3.55 \% & 0.20\\
No location specified & 26.47 \% & 28.80 \% & 0.92\\
No profile banner & 2.86 \% & 15.58 \% & 0.18\\
No URL in bio & 61.87 \% & 61.14 \% & 1.01\\
Customized profile theme & 17.67 \% & 33.14 \% & 0.53\\
Use the default profile image & 2.26 \% & 2.96 \% & 0.76\\
$||$Name$||$ < 3 & 4.40 \% & 2.56 \% & 1.72\\
$||$Name$||$ > 20 & 17.01 \% & 14.40 \% & 1.18\\
Have less than 10 tweets & 1.07 \% & 3.75 \% & 0.29\\
 Have less than 100 tweets & 5.06 \% & 10.45 \% & 0.48\\
 Created since 2019 & 92.81 \% & 52.66 \% & 1.76 \\
 Have verified badge & 1 & 2 & 0.50 \\[0.25ex]\hline
\end{tabular}
\caption{Analysis of outliers and individual user attributes. ``$||$Name$||<3$'' stands for the number of unique accounts that have less than 3 characters in their names. The number of unique users having a specific attribute is normalized by the total number of unique users in the corresponding class.}
\label{table:individual-attributes}
\end{table*}


%% file: detection.tex
\label{detection}
While there exist some studies on social and psychological aspects of doxing, an automated tool is still needed for the timely identification of such events with greater scalability. This is because once the private information is exposed to the public in a social network such as Twitter, it cannot be withdrawn again. Moreover, we have observed that many of the tweets in our filtered collection have been deleted, either by Twitter or the authors themselves. Additionally, some of the accounts have been suspended since then. Nevertheless, it is crucial to note that we have all those tweets and sensitive information collected offline using the streaming API before Twitter figures out there is a violation of its privacy terms and deletes the tweet or the author. In this paper, our goal is to automatically detect malicious purported sensitive information disclosure and doxing. We filter out tweets that do not contain private information, even if they talk about such information but the actual information (e.g., SSN) is not present. Furthermore, we discard self-disclosures and sensitive information that do not target any specific and uniquely identifiable person, making them not suspicious or harmful. Thus, the novelty of our approach is in discerning the difference between (1) nonsensitive and harmless information disclosures and self-disclosures (last row of Table~\ref{table:intentions}), and (2) doxing and malicious sensitive disclosures (all other rows).

\subsection{Feature Extraction}
We present and compare nine different detection approaches, as shown in Table~\ref{table:comparison-approach}, each of which uses a unique configuration and one or a combination of feature extraction methods as indicators of sensitive information. {\color{dgreen}We have selected these feature extraction methods and pretrained linguistic models due to their abundant applications in the literature for similar binary text classification tasks~\cite{xu2012learning, bellmore2015five, canfora2018nlp, umar2019detection, xu2019detecting, martin2021deep} and automated cyberbullying detection on Twitter~\cite{al2018optimized, banerjee2019detection, dadvar2020cyberbullying, kumar2020multi, torres2019cross, wrobel2019approaching, yimam2020exploring, srivastava2021role}.}

\paragraph{\textbf{String-matching heuristics}} Natural language patterns and string-matching have been used in various studies for identification of cyberbullying~\cite{xu2012learning, bellmore2015five} and privacy leaks~\cite{canfora2018nlp, umar2019detection, xu2019detecting}.
Our heuristics-based method searches for specific strings in tweets. We have identified a set of 29 phrases that based on our collected tweets, are more likely to indicate whether a tweet is doxing (positive rule) or non-doxing (negative rule). Examples of doxing indicators are tweets containing statements like ``\textit{your ip address is},'' ``\textit{your photos will be posted},'' or ``\textit{your ssn is}.'' We assume in the presence of sensitive information such as IP address or social security number, tweets containing these phrases are most likely intending to threaten a second party by revealing private information about her. We augment our phrase heuristics with some offensive and vulgar terms mostly used to offend victims in our tweets. These heuristics that are listed in Table~\ref{table:heuristics}, also include 24 \textit{invalid-looking} SSNs that seemed to be invalid to us and were mostly used for joking around or posting nonsensitive tweets. Table~\ref{table:invalid-ssns} shows possible figurative meanings\footnote{We have identified these digits and their possible figurative meanings by analyzing their usages in our tweet collection and searching for their common figurative meanings.} of some digits that can be found in our set of invalid-looking SSNs. All the negative rules (rules that label a tweet as negative) precede and overrule the positive ones. For instance, if a tweet contains an invalid-looking SSN and mentions the word ``dox,'' this approach labels the tweet as negative.


For IP address, we considered the string ``\textit{[Fail2Ban] POSTFIX-neelix}'' as an indicator of not being doxing because we found relatively many samples containing that string while the reported IP addresses were not associated with any specific identities and did not have any other information linked to them. Therefore, we did not believe they could be sensitive. A compound heuristic was when we observe the term ``user'' or \textit{@username} mentions and longitude and latitude values along with the IP address. We also considered the presence of the string ``\textit{you live in}'' along with an IP address as an indicator of doxing.

\begin{table*}[ht]
\centering
\begin{tabular}{|c||l|} 
 \hline
 Type of heuristic & Strings \\ [0.5ex] 
 \hline\hline
  & \textit{left me no choice, this you, deactivate luv, dox, cyberbully, loser, watch out,}\\Common& \textit{scare, dumb, your ip address is, your ssn is, blacks, black people, nigga, hate, }\\phrases& \textit{fuck, bitch, shut up, delete this or, delete the video or, warned you, delete the }\\& \textit{image or, death, troll, ass, shit, delete your video, i have your ip address, }\\&
  \textit{your photos will be posted}\\\hline 
  & \textit{111-11-1111, 222-22-2222, 333-33-3333, 444-44-4444, 555-55-5555,}\\Invalid-looking & \textit{666-66-6666, 777-77-7777, 888-88-8888, 999-99-9999, 123-45-6789, }\\SSNs& \textit{696-96-9696, 420-69-6969, 420-69-6669, 420-69-6666, 420-69-1488}\\& \textit{420-69-1337, 420-69-8008, 420-69-1312, 420-69-1313, 420-69-1234, }\\& \textit{420-69-2001, 420-69-1969, 420-69-1738, 078-05-1120}\\[0.25ex]\hline
\end{tabular}
\caption{Phrases and invalid-looking SSNs we used as our heuristics}
\label{table:heuristics}
\end{table*}

\begin{table*}[ht]
\centering
\begin{tabular}{|c||l|} 
 \hline
 Indicators & Possible figurative meaning \\ [0.5ex] 
 \hline\hline
  xxx-xx-xxxx &  All the digits are the same\\
  123-45-6789 & The sequence of all digits\\
  078-05-1120 & The SSN used and widely distributed by the Woolworth wallet manufacturer{\color{dgreen}.}\footnote{https://www.ssa.gov/history/ssn/misused.html}\\
  69 & A sex position\\
  420 & Slang for Marijuana and smoking pot\\
  666 & Known as the number of the beast or Devil's number\\
  1234 & The sequence of numbers (if used along with 420 and 69)\\
  1312 & Its digits represent the first three alphabet letters, ``ACAB,'' which is an\\& acronym used as a political slogan by those who are opposed to the police\\& and stands for ``All Cops Are Bastard''\\
  1313 & Consists of a pair of 13s which is sometimes referred to as the number of bad\\& (or good) luck or new beginnings\\
  1337 & Represents the term ``LEET''\\
  1488 & The 14-word slogan of white supremacists and 8 stands for `H', the \nth{8} letter \\& in alphabet, and 88 is a code representing the initials for ``Heil Hitler''\\
  1738 & Remy Martin Cognac\\
  1969 & A birth year in the \nth{20} century that ends with 69\\
  2001 & The year of the September 11 attacks\\
  8008 & A representation for the term ``BOOB''\\[0.25ex]\hline
\end{tabular}
\caption{Possible figurative meaning of specific digits that may be indications of invalid SSNs and false disclosures.}
\label{table:invalid-ssns}
\end{table*}

\paragraph{\textbf{One-hot encoded heuristics}}
Although we repeatedly observed some phrases and strings in doxing episodes, a simple string-matching technique is not capable of weighting each keyword and differentiating between more and less significant ones properly. To create fairer decision-making based on our heuristics, we built a classifier using heuristics in the form of one-hot encoded features. Each phrase or invalid-looking SSN in our set of heuristics is considered as a unique feature, and the presence (+1) or absence (-1) of each string in a tweet creates the feature vector corresponding to that tweet. Therefore, the number of features in this method is the same as the number of features in the string-matching feature extraction method. These feature vectors will be fed to a binary classifier to detect sensitive disclosure samples.

\paragraph{\textbf{Classical word embeddings}}
A more generalizable approach that does not rely on any category-specific heuristics is based on word embeddings. Word embeddings can be used for vectorization and mapping textual contents into vectors of numbers understandable by computers. Because training such models is usually highly time- and resource-consuming, typically a pretrained model is being used to find vectors associated with each word in the text. A common pretrained word embedding model that has been widely used in cyberbullying studies~\cite{al2018optimized, banerjee2019detection, dadvar2020cyberbullying} is GloVe~\cite{pennington2014glove}. Each word can be associated with a vector with different dimensions such that words with semantic or syntactic relations are in close vicinity~\cite{kumar2020multi}. 

We use the GloVe word embedding with 200 dimensions that is trained on a Twitter corpus of two billion tweets\footnote{https://nlp.stanford.edu/projects/glove} (it outperformed the similar lower-dimensional models in our analyses). This model has been trained on the same type of data that we are working with (tweet), and for this phase of our study, we did not have enough labeled samples to create an appropriate corpus and build a domain-specific model. Each tweet can be considered a set of vectors, each of which has 200 features and represents a single token. We take the average of all the vectors corresponding to tweet words and generate a single 200-dimensional feature vector per tweet.

\paragraph{\textbf{Contextualized string embeddings}}\label{string-embedding}
While classical word embedding models have been extensively studied and compared in the literature, they are not the best approach for vectorizing tweets that may consist of several sentences and misspelled words. Contextualized string embedding is a type of contextualized character-level embedding which captures word semantics in the context and produces different embeddings for polysemous words based on their usage~\cite{akbik2018coling}. It models words and their contexts as a sequence of characters and can perform better on misspelled words that are common on social media posts. 

{\color{dgreen}Prior studies show its appealing performance for various text classification tasks such as detection of cyberbullying~\cite{wrobel2019approaching} and crisis-related~\cite{torres2019cross} tweets, and identification of hate speech in Hinglish~\cite{srivastava2021role}.} We use the pretrained forward language model with 2048 features from the Flair contextualized string embeddings~\cite{akbik2019flair}. We also use document pool embeddings from the Flair framework to stack embedding models~\cite{torres2019cross, yimam2020exploring} and obtain a unique feature vector for the whole tweet because the word and contextualized string embeddings provide word-level and character-level features respectively. Document pool embedding calculates an average of all embeddings in a sentence.

\subsection{Classification}
In~\S\ref{sec:analyses}, we propose different approaches to use these feature extraction methods and perform a binary classification task, using a classifier, string-matching heuristics, or both. Table~\ref{table:configurations} presents the type and number of features we use for each of the classifiers, as well as the size of samples for each of the two classes and the train and test portions of the data. Except for the first approach that is solely based on string-matching and do not include any learning phase (uses the whole data for testing), we use stratified 10-fold cross-validation to split our samples with a consistent ratio between both classes and make sure our test data is representative enough, and the classifier is not overfitted or biased. We use the LinearSVC implementation of Support Vector Machines (SVM) from the Python Scikit-Learn library\footnote{https://scikit-learn.org} with its default argument values as our learner and classifier in all our classification tasks. 

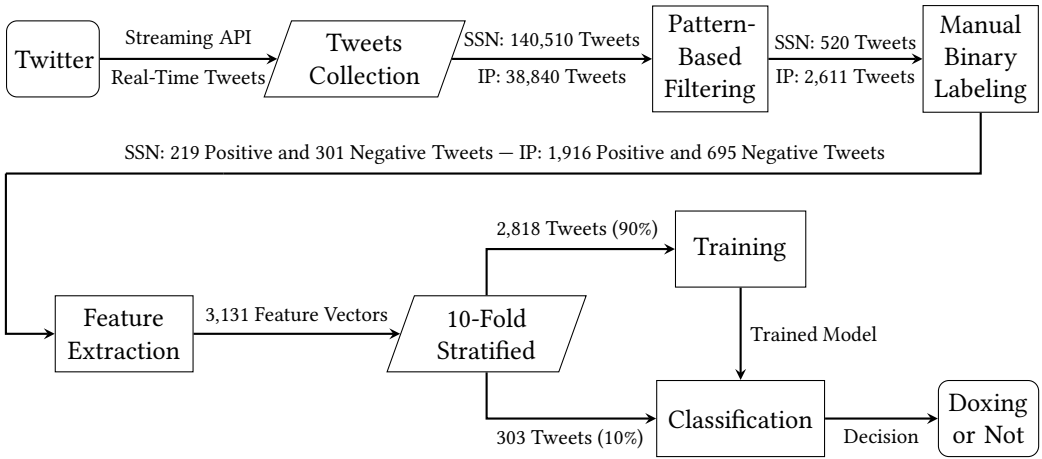
\begin{figure}[!h]
\begin{tikzpicture}[node distance=2.1cm]
    \node (twitter) 
    [startstop, minimum width=0.02cm, text width=1cm, text centered] 
    {Twitter};
    \node (collection) 
    [io, right of=twitter, minimum width=0.4cm, text width=1.5cm, text centered, xshift=2.05cm] 
    {Tweets Collection};
    \node (filter) 
    [process, right of=collection, minimum width=1cm, text width=1.3cm, text centered, xshift=2.5cm] 
    {Pattern-Based Filtering};
    
    \node (labeling) 
    [process, right of=filter, minimum width=1cm, text width=1.3cm, text centered, xshift=1.5cm] 
    {Manual Binary Labeling};
    
    \node (invisible-l) [coordinate,below right = 1.0cm and -1.25cm of twitter] {};
    
    \node (invisible-r) [coordinate, right = 2.9cm of invisible-l] {};
    
    \node (extraction) 
    [process, below right = 2.6cm and -0.6cm of twitter, minimum width=1cm, text width=1.6cm, text centered] 
    {Feature Extraction};
    
    \node (vectors) 
    [io, right = 2.75cm of extraction, minimum width=1cm, text width=1.5cm, text centered]
    {10-Fold Stratified};
    
    \node (training) 
    [process, above right= 0.1cm and 2.0cm of vectors, minimum width=1cm, text width=1.5cm, text centered]
    {Training};
    
    \node (classification) 
    [process, below right= 0.1cm and 1.75cm of vectors, minimum width=0.8cm, text width=2cm, text centered] 
    {Classification};
    
    \node (decision) 
    [startstop, right of=classification, minimum width=0.02cm, text width=1.1cm, text centered, xshift=1.2cm] 
    {Doxing or Not};

    \draw [arrow] (twitter) -- node[anchor=south] {\footnotesize Streaming API} node[anchor=north] {\footnotesize Real-Time Tweets} (collection);
    
    \draw [arrow] (collection) -- node[anchor=south] {\footnotesize SSN: 140,510 Tweets} node[anchor=north] {\footnotesize IP: 38,840 Tweets} (filter);
    
    \draw [arrow] (filter) -- node[anchor=south] {\footnotesize SSN: 520 Tweets} node[anchor=north] {\footnotesize IP: 2,611 Tweets} (labeling);
    
    \draw [thick] (labeling) |- (invisible-r);
    
    \draw [thick] (invisible-r) -- node[anchor=south west]{\footnotesize SSN: 219 Positive and 301 Negative Tweets — IP: 1,916 Positive and 695 Negative Tweets} node[anchor=north west]{} (invisible-l);
    
    \draw [arrow] (invisible-l) |- (extraction);
    
    \draw [arrow] (extraction) -- node[anchor= south] {\footnotesize 3,131 Feature Vectors} node[anchor=north] {} (vectors);
    
    \draw [arrow] (vectors) |- node[anchor=south west] {\footnotesize 2,818 Tweets (90\%)} (training);
    
    \draw [arrow] (vectors) |- node[anchor=north west] {\footnotesize 303 Tweets (10\%)} (classification);
    
    \draw [arrow] (training) -- node[anchor=west] {\footnotesize Trained Model} (classification);
    
    \draw [arrow] (classification) -- node[anchor=north] {\footnotesize Decision} (decision);
\end{tikzpicture}
\caption{Overview of our dataset creation and enrichment pipeline, combined with our experimental setup. Positive tweets are the ones that are doxing or malicious disclosures and negative tweets are the non-doxing tweets and self-disclosures.}
\label{fig:pipeline}
\end{figure}

Furthermore, we have performed our classifications with and without retweets and those tweets containing quoted tweets. However, we have kept them in our analyses because they are present in practice and yield better performance. {\color{dgreen} Note that we aim to identify sensitive disclosures of SSN and physical location obtained by IP address. Our approach does not aim to detect all types of sensitive private information disclosures. Therefore, we only annotate and use our filtered samples for training and testing. Our pattern-matching filter, introduced in \S\ref{sec:init-filtering}, however, can be used to exclude tweets that do not contain digits for SSNs and IP addresses and feed the remaining potentially sensitive tweets to our system. \Cref{fig:pipeline} illustrates an overview of our tweet collection, dataset creation, modeling, and automated detection of doxing samples, as well as the number of tweets used in each step.}

%% file: analyses.tex
\label{sec:analyses}
We use our string-matching feature extraction as the baseline and compare it with eight other methods and configurations to detect second- and third-party private information disclosures. As it is shown in Table~\ref{table:comparison-approach}, the baseline approach is capable of detecting one-third of episodes and provides a high ratio of detection for the negative (non-doxing and unmalicious) class. This is because even after using an exhaustive set of heuristics indicating positive and negative classes, this approach can only make decisions for the tweets that either do or do not contain specific strings. However, it cannot be appropriately generalized to detect every new tweet automatically. Note that the presented metrics in Table~\ref{table:comparison-approach} reflect the performance for detecting the positive class (doxing and malicious disclosures) as we believe flagging the positive samples is more critical and sensitive compared to the negative samples. We use the default negative label if a tweet does not match any of our positive or negative rules. Therefore, some tweets are identified as negative only because they have not matched any rules. Moreover, this approach requires contextually- and linguistically-specific lexicons that can be continuously updated~\cite{tawesu_tweets_2019}. While this approach considers tweets containing any of the SSNs listed in Table~\ref{table:heuristics} as negative, the heuristics that we use for IP addresses consist of both positive and negative rules.

In contrast to the simple string-matching, one-hot encoded heuristics (1-HotEH) led to a learner that most often classifies the tweets as positive, and its performance is not desirable for detecting the negative samples. In addition to the initial heuristics, we experimentally added one-hot encoded English pronouns to these features to augment our set of features and make them more representative. However, they did not make any considerable changes to our results. Although the number of features in this approach was less than the number of training samples (even with pronouns), SVM could not converge in some of the rotations of our cross-validation which may make the predictions unreliable. 

Table~\ref{table:comparison-approach} also shows how the combination of these two methods performs (1-HotEH\_Heuristics). The string-matching heuristics precede the classifier and overrule for any combination of classifiers and heuristics. We disregard the classifier's decision if our heuristics match a tweet. As it can be seen, there is only a subtle improvement in metrics compared to each of the two methods individually. The next approach utilizes the GloVe Twitter model (Mean\_GloVe\_Twitter). To combine the feature vectors that the model gives us per word (if the word exists in the embedding's dictionary), we take the average of all the vectors we extracted for the tweet's words. As it is shown in Table~\ref{table:comparison-approach}, this approach is more reliable and robust compared to the previous ones.

In the next approaches, we use document pool embedding (DP) to combine word embeddings and contextual string embeddings of each tweet. The next model that we use is the GloVe Wikipedia model (DP\_GloVe\_Wiki) that outperforms the previous GloVe-based approach. We obtained the best and most robust overall performance on all our classification metrics using the Flair forward model (DP\_FlairFW). Out of the 520 tweets we had for the SSN subcategory, we found 102 tweets (20\%) that their SSNs were on our invalid-looking SSNs list. Therefore, we performed a separate experiment by removing those samples from our dataset to illustrate the difference they may make in our detection performance (DP\_FlairFW\_Cleaned). These 102 tweets consist of 99 negative and three positive samples. While we believe these SSNs are invalid or fake, we labeled them, disregarding this assumption and behaved them as other SSNs. Although we ran all our experiments with and without these samples, we have only reported one instance in Table~\ref{table:comparison-approach} for brevity because they did not make a notable difference in any of our analyses. We even removed them in the first experiments of heuristics and one-hot encoded heuristics, but again, the performance did not change considerably. 

In the next phase, we combined DP\_FlairFW with our overruling heuristics, but it did not make a drastic difference and only improved the recall while impairing other significant metrics. Our last analysis uses a document pool embedding of both Flair forward and GloVe models with the total features of 2148 per tweet, but it almost did not make any difference compared to DP\_FlairFW. We also stacked Flair's forward and backward pre-trained models as they were stacked in prior studies for bot detection on Twitter~\cite{martin2021deep} but did not observe improvements in the performance.

To illustrate the statistical significance of the outperformance achieved by our proposed approach over our baseline 1-HotEH approach, we have used \textit{five 2-fold cross-validation (5x2cv) paired t-test}~\cite{dietterich1998approximate}. Such a selection is shown to achieve the most possible independence between train and test samples for hypothesis testing. The 2-fold cross-validation makes sure that in each trial (out of the total five trials suggested by Dietterich~\cite{dietterich1998approximate}), distinct and non-overlapping samples are being used for training and testing the classifiers. Moreover, the trials need to be repeated five times so that various samples have been used to train and test the classifiers and are not overfitted. We obtained the \textit{t-value} of 166.06 for comparing the 1-HotEH and DP\_GloVe\_Wiki approaches. 

According to t-value tables for one-sided tests and the degree of freedom for the resulting Student's t-distribution ($df=5$), this t-value corresponds to a \textit{p-value} less than 0.00005, which indicates the significance of this improvement with the confidence level of 0.9995. Similarly, the \textit{t-value} we calculated for improvement by DP\_FlairFW over 1-HotEH was 81.03 which provides a \textit{p-value} in the same range. While we have presented the results of our analyses of various feature extraction methods, the main goal of this paper is to detect doxing and malicious private information disclosures on Twitter which has been largely under-investigated. Therefore, we have evaluated our automated detection approach by comparing various implementations of it using similar and potentially-effective feature extraction methods from the literature. Our goal was to illustrate the practicality and robustness of the approach in detecting personal information disclosures and differentiating the malicious and doxing episodes from self-disclosures and nonsensitive disclosures.

\begin{table*}[ht]
\centering
\begin{tabular}{|l||c|c|c|c|c|c|} 
 \hline
 Method & Features & Feat. & Pos. & Neg. & Train & Test\\ [0.5ex] 
 \hline\hline
 Heuristics & Table~\ref{table:heuristics} strings and & N/A & 2135 & 996 & N/A & 3131\\& IP address heuristics &&&&&\\\hline 
 1-HotEH & One-hot encoded & 67 & 2135 & 996 & 2818 & 313\\& strings from Table~\ref{table:heuristics} &&&&&\\\hline 
 1-HotEH\_Heuristics & One-hot encoded strings, & 67 & 2135 & 996 & 2818 & 313\\& overruling heuristics &&&&&\\\hline 
 Mean\_GloVe\_Twitter & Average of GloVe Twitter & 200 & 2135 & 996 & 2818 & 313\\& model word embeddings &&&&&\\\hline 
 DP\_GloVe\_Wiki & Document pool embedding & 100 & 2135 & 996 & 2818 & 313\\& of GloVe Wikipedia model &&&&&\\\hline
 DP\_FlairFW & Document pool embedding & 2048 & 2135 & 996 & 2818 & 313 \\& of Flair news forward model &&&&&\\\hline
 DP\_FlairFW\_Cleaned & Document pool embedding & 2048 & 2132 & 897 & 2726 & 303\\& of Flair news forward model &&&&&\\\hline
 & Document pool embedding &&&&&\\ DP\_FlairFW\_Heuristics & of Flair news forward model, & 2048 & 2135 & 996 & 2818 & 313\\& overruling heuristics &&&&&\\\hline
 DP\_FlairFW\_GloVe\_Wiki & Document pool embedding & 2148 & 2135 & 996 & 2818 & 313\\& of Flair and GloVe models &&&&&\\[0.25ex] 
 \hline
\end{tabular}
\caption{Configurations used for different detection approaches. ``Feat.,'' ``Pos.,'' and ``Neg.'' stand for ``Features,'' ``Positive samples,'' and ``Negative samples'' respectively. Note that the train and test sizes are the average sizes per fold in a 10-fold stratified cross validation, except for the first row which does not require any training. Also, the cleaned dataset has 94 tweets less than others in totals which are the invalid-looking SSNs that are removed. Furthermore, the number of features only represents the features used for an automated classification task and does not include the strings used as our heuristics.}
\label{table:configurations}
\end{table*}

\begin{table*}[ht]
\centering
\begin{tabular}{|l||c|c|c|c|c|c|c|c|} 
 \hline
 Method & TPR & TNR & FPR & FNR & Acc. \% & Prec. \% & Rec. \% & F1 \% \\ [0.5ex] 
 \hline\hline
 Heuristics & 0.20 & 0.62 & 0.38 & 0.80 & 33.47 & 53.22 & 20.14 & 29.22\\\hline 
 1-HotEH & 0.99 & 0.11 & 0.90 & 0.01 & 71.10 & 70.42 & 99.34 & 82.42\\\hline 
 1-HotEH\_Heuristics & 1.00 & 0.10 & 0.90 & 0.00 & 71.19 & 70.35 & 99.81 & 82.53\\\hline 
 Mean\_GloVe\_Twitter & 0.97 & 0.92 & 0.08 & 0.03 & 95.37 & 96.19 & 97.05 & 96.62\\\hline 
 DP\_GloVe\_Wiki & 0.97 & 0.93 & 0.07 & 0.03 & 95.46 & 96.76 & 96.58 & 96.67\\\hline
 DP\_FlairFW & 0.94 & 0.87 & 0.13 & 0.06 & 91.25 & 92.58 & 93.64 & 93.11\\\hline
 DP\_FlairFW\_Cleaned & 0.97 & 0.96 & 0.04 & 0.03 & 96.86 & 98.16 & 97.37 & 97.76\\\hline
 DP\_FlairFW\_Heuristics & 0.98 & 0.90 & 0.10 & 0.03 & 95.05 & 95.37 & 97.47 & 96.41\\\hline
 DP\_FlairFW\_GloVe\_Wiki & 0.97 & 0.95 & 0.05 & 0.03 & 96.61 & 97.74 & 97.28 & 97.51\\ [0.25ex] 
 \hline
\end{tabular}
\caption{Comparison of different detection approaches. ``Acc.,'' ``Prec.,'' ``Rec.,'' and ``F1'' stand for ``Accuracy,'' ``Precision,'' ``Recall,'' and ``F1-score'' respectively.}
\label{table:comparison-approach}
\end{table*}

%% file: limitations.tex
This paper presents a  comprehensive study of the doxing phenomenon, with insights into potential intentions behind sensitive disclosures. It also proposes an automated approach to detect two important and highly sensitive information disclosures. 
Here, we discuss some limitations of our work and findings. First, although we have collected and analyzed a broad category of sensitive information, our labeling,   analyses of intentions, and detection performance is limited to only two highly sensitive types of private information. Therefore, it is likely and expected that the proposed approach misses doxing incidents that do not contain these types of information.

Moreover, as shown by the calculated {\color{dgreen}Fleiss'} Kappa coefficient, the labeling task is to some extent subjective, and different people and annotators may have conflicting perceptions of sensitive information disclosures and their maliciousness. For instance, in case of disclosing information as a result of feeling desperate, we were skeptical about some \textit{seemingly self-disclosures} and whether they try to impersonate another individual and reveal the {\color{dgreen}victim's} real sensitive information in a way that does not look highly suspicious to Twitter or individuals. Furthermore, we use a set of keywords that we found to be more likely included in the tweets that reveal such sensitive information while avoiding high false positives and collecting more potentially-doxing samples, and creating a more balanced dataset.

Although we do not claim or prove that these keywords are the most effective and efficient keywords for collecting corresponding sensitive information, the distribution of positive and negative classes for each subcategory in Table~\ref{table:keywords} depicts the effectiveness of the chosen keywords. However, identifying such keywords for other categories of sensitive information would require an initial manual inspection of tweets containing those types of information and iteratively augmenting the set of keywords. We have also collected our dataset during several months and with interruptions to make sure it contains different information disclosure events in which a group of users may rapidly tweet or retweet sensitive information during a short period. Yet, our dataset contains duplicate sensitive information (e.g., the same SSN has been used in multiple tweets) either in the same context (e.g., retweet) or divergent contexts {\color{dgreen}that may be even different from the original disclosure incident and unrelated to the actual owner of such private information (e.g., an IP address that is disclosed in one event and might be valid, is used to falsely dox or scare another person)}. While this leads to a realistic sample, the resulting data may be noisy and affect the performance of our classifiers. We also did not perform any URL analysis, however, doxed information can be posted to anonymous content and document sharing platforms such as \textit{4chan.org}, \textit{doxbin.org}, and \textit{pastebin.com}, and reshared on Twitter as a generic URL and evade our text-based classifiers.

{\color{dgreen}   For obvious ethical reasons, we could not present the exact tweet contents and reveal private information that exists in our dataset. We have redacted personal information such as people's names, screen names of users, SSNs, IP addresses, and dates from the examples that are presented in this paper.} And finally, while we have discarded some patterns and sensitive information that are evidently fake, we cannot verify the purported information as we do not have access to all the (potential) victims of the disclosures. Therefore, we cannot differentiate between falsely purported malicious disclosures and doxing.\\

%% file: conclusion.tex
{\color{dgreen}Doxing can lead to physical threats and detrimental effects on people's lives and professional careers by exposing their sensitive private information.} In this paper, we have created a manually-labeled dataset of tweets for doxing detection based on which we provided a deep insight into sensitive information disclosure motivations. We also proposed an automated detection of doxing and malicious information disclosures on Twitter. We differentiated these two from nonsensitive disclosures and self-disclosures as the first two are more important to be identified. They can have higher risks, especially if victims are not aware of the incidents and take timely measures (e.g., reporting a tweet) to protect themselves. Such an automated approach is vital because once the private information is exposed to the public, it cannot be easily withdrawn, if at all. 

We have also presented examples of suspicious tweets and accounts that have been removed or suspended later, highlighting that using a standard and free account, one can access sensitive private information without any repercussion from Twitter. We have compared several implementations of our proposed detection approach using various feature extraction methods. While our approach does not require category-based features and separate classifiers, it can detect disclosures of SSNs and IP addresses associated with specific second- and third-parties by specifying their other personally identifiable information such as their usernames or physical addresses. We differentiated between purported sharing of sensitive private information, malicious purported sharing of private information, and actual doxing episodes. 

While disclosures of SSNs and IP addresses can be very harmful, future studies can be performed using a similar approach on a larger labeled dataset collected using keywords associated with more categories of sensitive information to analyze its detection performance in a more diverse set of tweets. In practice, Twitter may use user reports to create such a dataset. However, crowdsourcing platforms can also be used to label a larger dataset by multiple annotators. Other features such as URLs, user attributes, and historical tweets can be studied on such a diverse dataset and potentially be incorporated to build an ensemble learner for automated detection of various types of doxed information. Moreover, none of the existing automated detection approaches for cyberbullying and doxing can completely identify these incidents. Therefore, user studies can be conducted to assess content moderation strategies and user experiences after augmenting the existing policies\footnote{https://help.twitter.com/en/safety-and-security/report-abusive-behavior} by implementing and enforcing content removals or various levels of protection against sharing such contents.

%% file: appendix.tex
\label{sec:appendix}
{\color{dgreen} Below we report the instructions and rules we provided to our annotators to make sure they have enough background information and are aware of requirements for labeling each tweet as being doxing or not:}\\

\begin{quote}

We aim to find the tweets in which private and sensitive information about \nth{2} or \nth{3} parties (not self-disclosures) is being disclosed without their consent (doxing), either intentionally or unintentionally, maliciously or  benignly.\\

\begin{itemize}
    \item Please use the column “doxing” in the sheet to mark each tweet as TRUE/T or FALSE/F. 
    \item You may also get help from the quoted status column which contains the quoted tweet, if there is any, to better understand the context. 
    \item If there are multiple tweets from the same user, you should not look up authors and make decisions based on their history. But if they have similar contents, you can take into account content similarity in your decision.
    \item Context of a doxing event obtained from other tweets (e.g., similar SSN used) can be considered in decision making.
\end{itemize}

\begin{table*}[ht]
\centering
\begin{tabular}{|l||l|} 
 \hline
 \textbf{Doxing (TRUE)} & \textbf{Non-doxing (FALSE)} \\ [0.35ex]\hline 
 Sensitive disclosure about \nth{2}/\nth{3} parties& \\(info is connected to victim’s identity by quoting & Self-disclosure\\ them, mentioning their full names, usernames, & \\ etc. regardless of their potential intentions.&\\\hline 
 Promoting doxed info (report/reply/quote of a & Does not target any\\tweet that contains doxing; even by the victim)& specific/unique identity/person\\\hline 
 Can be used to uniquely identify or physically& \\ locate a person &\\\hline
 There is no direction mention of other identities, &\\but there are indicators (e.g., your SSN is …)&\\[0.25ex]\hline
\end{tabular}
\caption{Specific rules and criteria we provided to our annotators for labeling a sample of 100 tweets from the dataset to calculate the inter-annotator reliability and agreement.}
\label{table:instructions}
\end{table*}

\end{quote}